\documentclass{elsart}
\usepackage{graphicx}
\usepackage{amssymb}
\begin{document}
\begin{frontmatter}
\title{Structure and evolution of the world trade network}
\author[1,2]{Diego Garlaschelli}
\and
\author[2,3]{Maria I. Loffredo}
\address[1]{Dipartimento di Fisica, Universit\`a di Siena, Via Roma 56, 53100 Siena ITALY}
\address[2]{INFM UdR Siena, Via Roma 56, 53100 Siena ITALY}
\address[3]{Dipartimento di Scienze Matematiche ed Informatiche, Universit\`a di Siena, \\Pian dei Mantellini 44, 53100 Siena ITALY}
\begin{abstract}
The \emph{World Trade Web} (WTW), the network defined by the international import/export trade relationships, has been recently shown to display some important topological properties which are tightly related to the Gross Domestic Product of world countries. While our previous analysis focused on the static, undirected version of the WTW, here we address its full evolving, directed description. This is accomplished by exploiting the peculiar reciprocity structure of the WTW to recover the directed nature of international trade channels, and by studying the temporal dependence of the parameters describing the WTW topology.
\end{abstract}
\date{9 September 2004}
\end{frontmatter}

\section{Introduction}
The internal structure of many biological, social and economic systems displays complex topological properties such as clustering, a scale-free degree distribution and the presence of degree correlations, which are not reproduced by simple random graph models \cite{albert}. 
The global trade activity is one example of a large-scale system whose internal structure can be represented and modeled by a graph \cite{wtw,mywtw}. In the most general case, from publicly available annual trade data \cite{data} it is possible to define the evolving \emph{World Trade Web} (WTW in the following), where each world country is represented by a \emph{vertex} and the flow of money between two trading countries is represented by a \emph{directed link} between them: if $i$ imports some good from country $j$ during the year $t$, then a link from $i$ to $j$ is drawn in the $t$-th snapshot of the graph, corresponding to a nonzero entry $a_{ji}(t)=1$ in the corresponding adjacency matrix. Otherwise, no link is drawn from $i$ to $j$ and $a_{ji}(t)=0$. Note that the direction of the link always follows that of the wealth flow, since exported (imported) goods correspond to wealth flowing in (out).
In such a description, if $N(t)$ denotes the number of world countries during year $t$, the \emph{in-degree} $k^{in}_i(t)=\sum_{j=1}^{N(t)}a_{ij}(t)$ and the \emph{out-degree} $k^{out}_i(t)=\sum_{j=1}^{N(t)}a_{ji}(t)$ of a country $i$ correspond to the number of countries importing from and exporting to $i$ respectively. 

The structure of the WTW continuously changes in time due to variations in the number $N(t)$ of world countries and to the continuous rearrangement of its connections, and hence of $a_{ij}(t)$. 
While previous studies focused on the undirected description of a single snapshot of the WTW \cite{wtw,mywtw}, in the present work we study the properties of the WTW as a directed and evolving network and we confirm that its topology always displays a peculiar dependence on the Gross Domestic Product of world countries, in excellent accordance with the so-called \emph{fitness model} \cite{fitness}. We first review the previous results and then extend them by reporting the temporal evolution of the relevant parameters.
Our analysis is based on a comprehensive dataset \cite{data} reporting the annual trade activity between all world countries together with the annual values of their total Gross Domestic Product during  the period 1950-1996.

\section{Topology of the WTW}
Two recent papers document some interesting static topological properties of the WTW viewed as an undirected network \cite{wtw,mywtw}. In such a representation, each pair of vertices $i,j$ is considered connected if $i$ and $j$ trade in at least one direction. Before reviewing these results, we first give the explicit expressions linking the directed and undirected versions of a graph.
From a graph-theoretic point of view, if $a_{ij}$ is the adjacency matrix of a directed graph, then the undirected version of the same graph is described by the symmetric adjacency matrix ${b_{ij}\equiv a_{ij}+a_{ji}-a_{ij}a_{ji}}$. Correspondingly, the quantities  ${k^{in}_i=\sum_j a_{ij}}$ and ${k^{out}_i=\sum_j a_{ji}}$ computed on the directed graph and the `undirected' degree ${k_i=\sum_j b_{ij}}$ computed on its undirected version are related through the following equation:
\begin{equation}\label{eq_und}
k_i=k^{in}_i+k^{out}_i-k^{\leftrightarrow}_i
\end{equation}
where we have introduced the \emph{reciprocal degree} \cite{reciprocity} ${k^\leftrightarrow_i\equiv\sum_j a_{ij}a_{ji}}$, defined as the number of neighbours of $i$ with double (reciprocal) connections to and from $i$. 
\begin{figure}[h]
\begin{center}
\includegraphics[width=.54\textwidth]{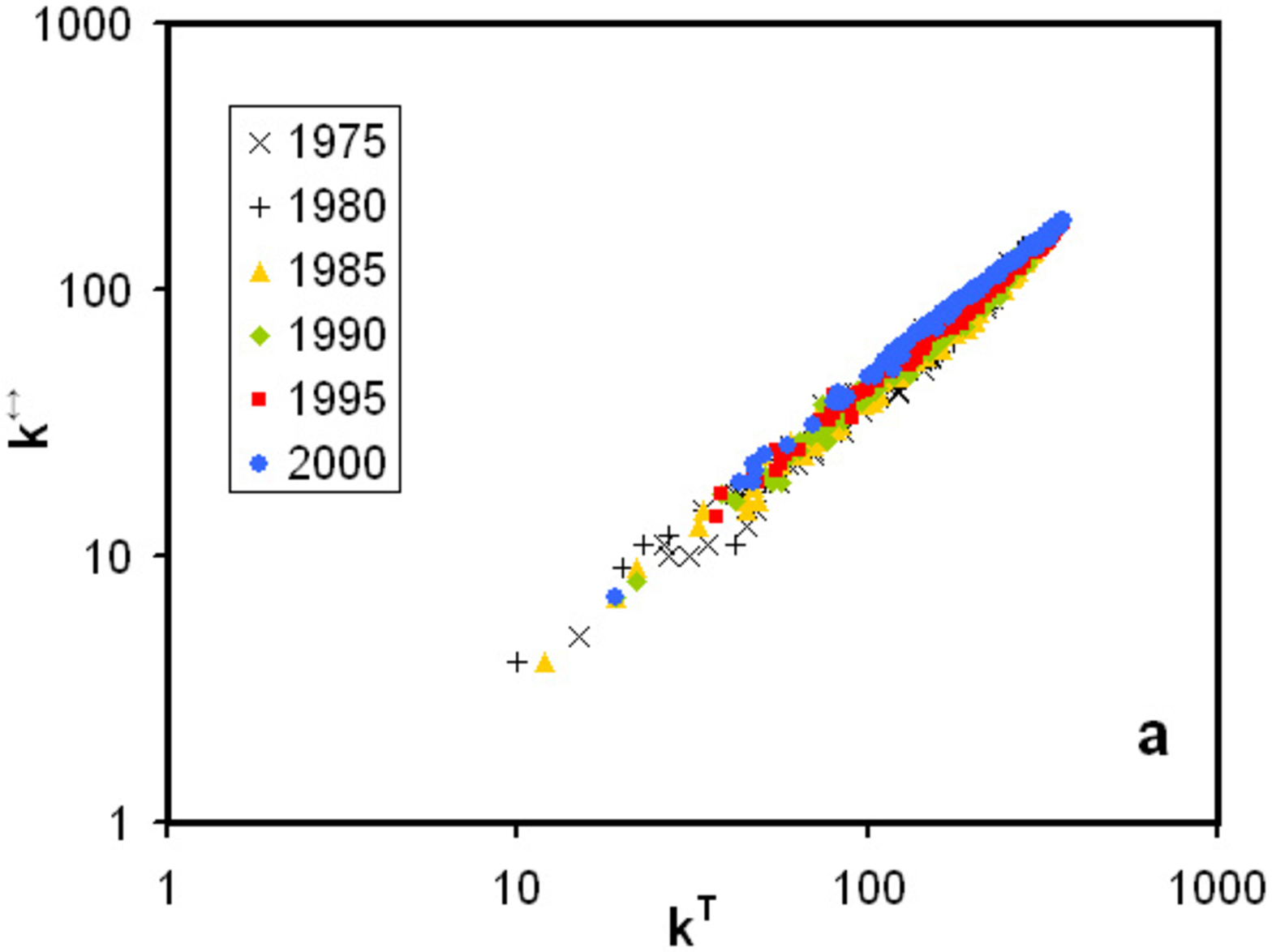}
\includegraphics[width=.45\textwidth]{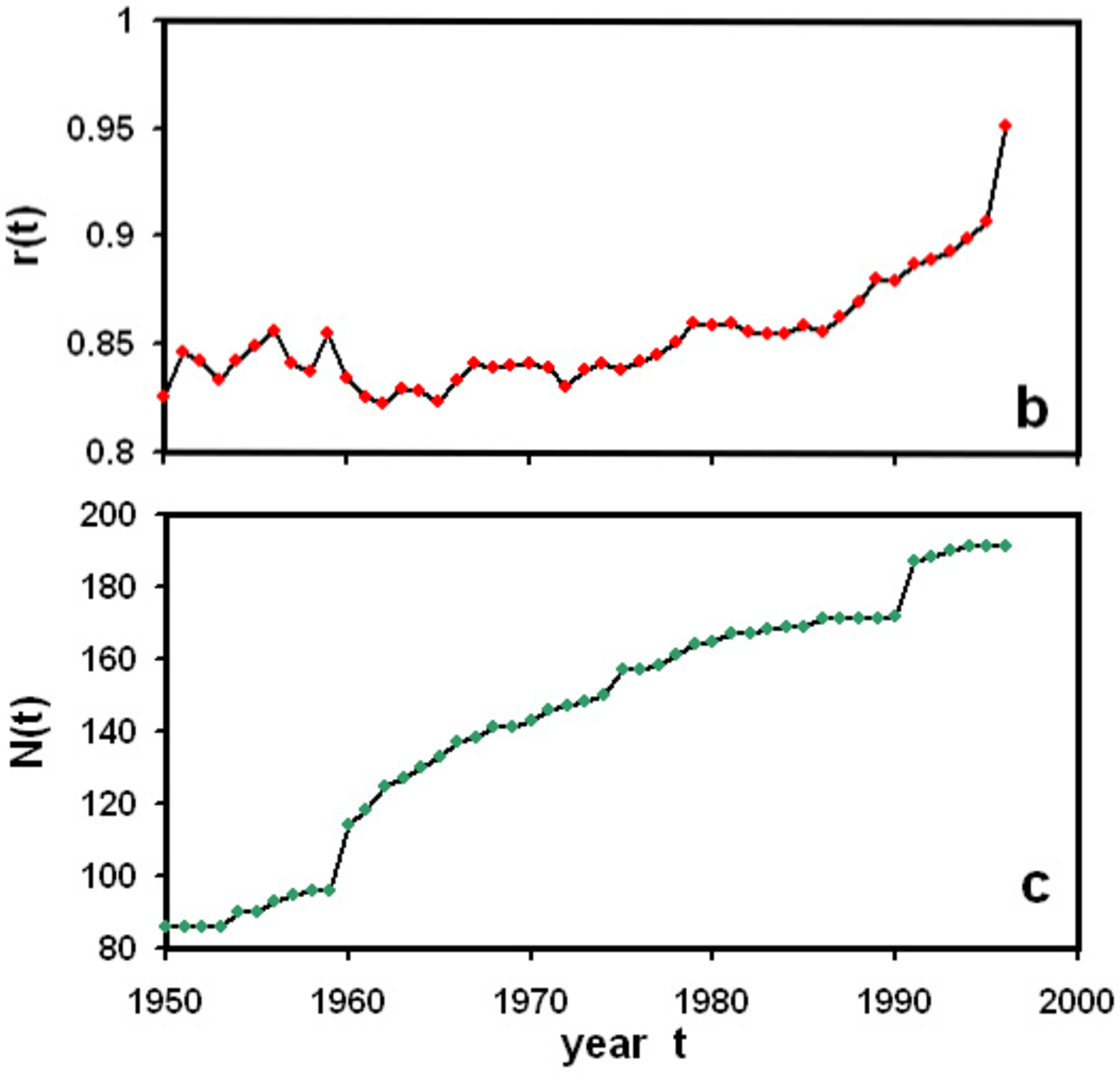}
\end{center}
\caption[]{Properties of the WTW. {\textbf a)} Reciprocal ($k^\leftrightarrow$) versus total ($k^T$) degree for various timesteps. {\textbf b)} Time evolution of the reciprocity $r(t)$. {\textbf c)} Time evolution of the number of world countries $N(t)$.}
\label{fig1}
\end{figure}
While in principle there is an obvious loss of information in using the undirected version of a directed graph, in a recent paper \cite{reciprocity} we showed that for the particular case of the WTW there are two important empirical properties that allow to recover the directed character from its undirected description. 
Firstly, one always observes that approximately ${k_i^{in}(t)\approx k_i^{out}(t)\quad \forall i,t}$. Secondly (see fig.\ref{fig1}a), a linear relation holds between $k_i^{\leftrightarrow}(t)$ and the \emph{total degree} ${k^T_i(t)\equiv k_i^{in}(t)+k_i^{out}(t)}$ \cite{reciprocity}:
\begin{equation}\label{eq_both}
k^\leftrightarrow_i(t)\approx \frac{r(t)}{2}k^T_i(t)
\end{equation}
where $r(t)$ is the \emph{reciprocity}, defined as the ratio of the number $L^\leftrightarrow(t)$ of links pointing in both directions to the total number of links $L(t)$:
\begin{equation}
r(t)\equiv\frac{L^\leftrightarrow(t)}{L(t)}
\end{equation}
The annual values of $r(t)$ for the 1950-1996 time interval are shown in fig.\ref{fig1}b. 
The above results allow to obtain the properties of the WTW, viewed as a directed graph, from its undirected version. For instance, eqs. (\ref{eq_und},\ref{eq_both}) imply
\begin{equation}
k^{in}_i(t)\approx k^{out}_i(t)\approx \frac{k^T_i(t)}{2}\approx\frac{k_i(t)}{2-r(t)}
\end{equation}
and a relation between $L$ and the number $L^u$ of links in the undirected network:
\begin{equation}
L(t)=\sum_{i=1}^{N(t)} k^{in}_i(t)=\frac{1}{2-r(t)}\sum_{i=1}^{N(t)}{k_i(t)}=\frac{2}{2-r(t)}L^u(t)
\end{equation}

As a consequence, in the following we can simply address the topology of the undirected WTW and then recover, at any timestep $t$, its full directed description through the value of $r(t)$. For simplicity, we now review the results corresponding to a single snapshot (the year 1995) of the undirected WTW \cite{mywtw}. An important result that we report here is that the same qualitative trends are observed for each year in the database (1950-1996) and can be described quantitatively by plotting the temporal evolution of the relevant parameters. For instance, in fig.\ref{fig1}c we show the evolution of the number of countries $N(t)$ during such time interval. For the 1995 data, the number of countries equals $N=191$.

\begin{figure}[h]
\begin{center}
\includegraphics[width=.52\textwidth]{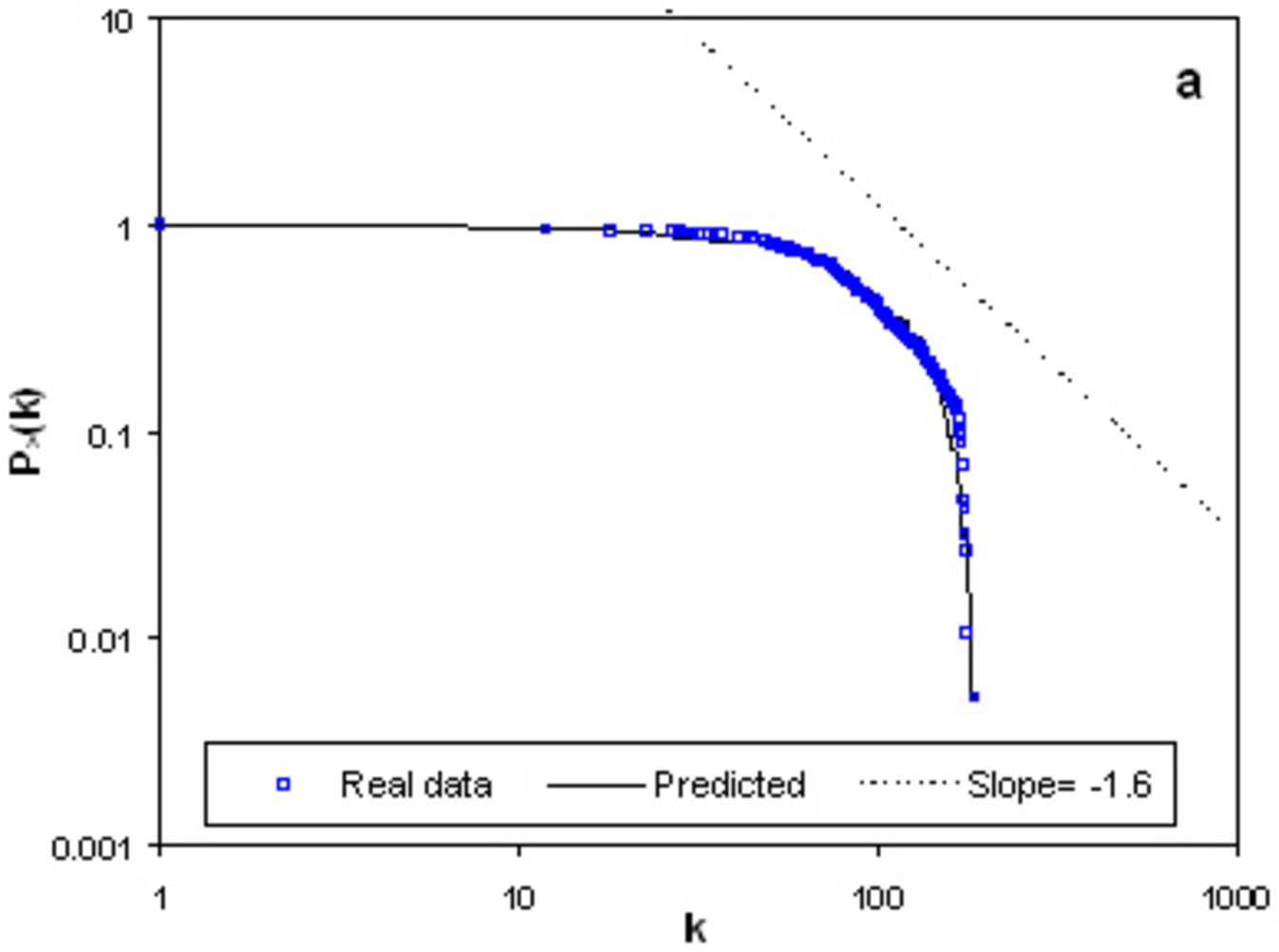}
\includegraphics[width=.36\textwidth]{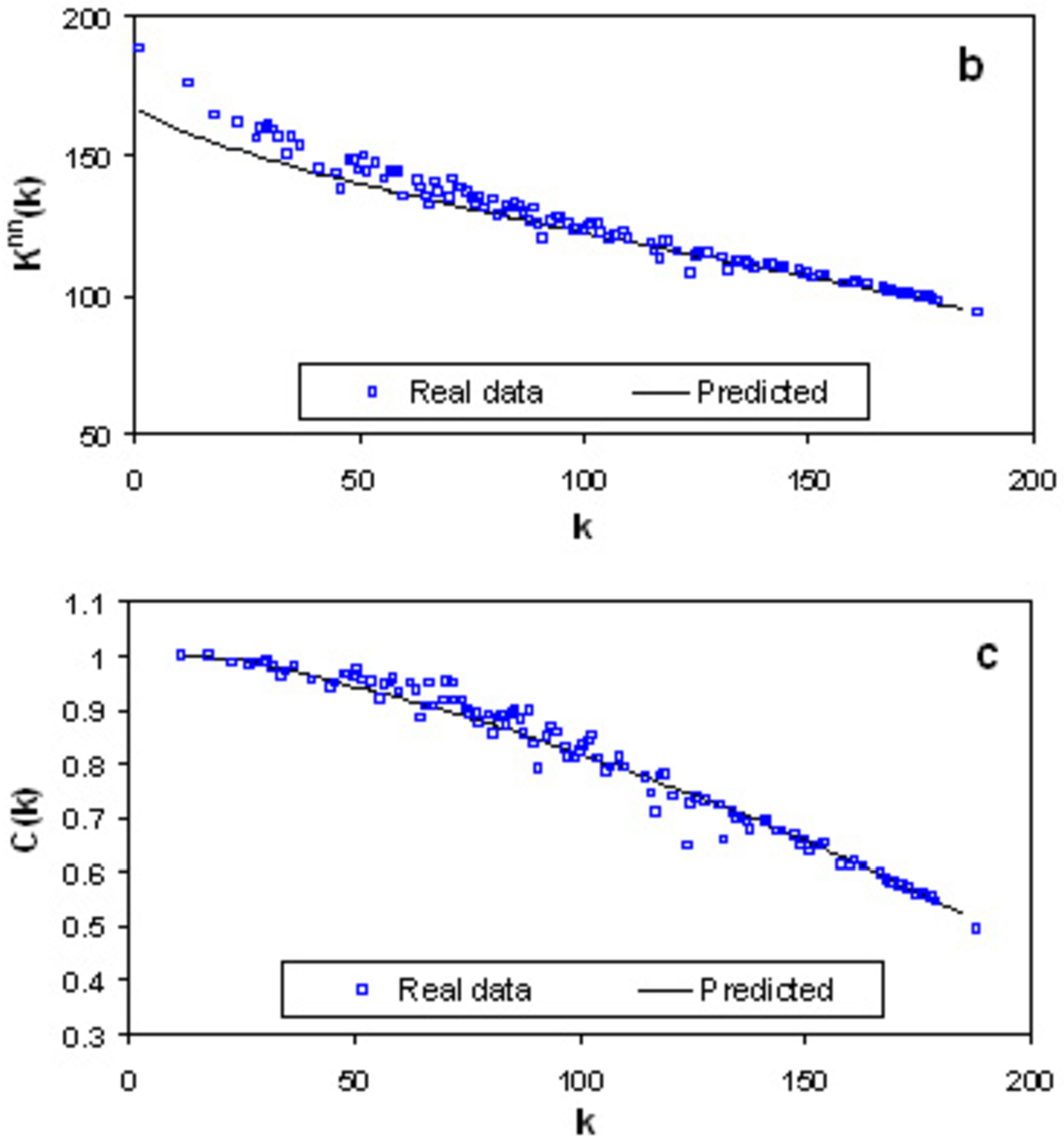}
\end{center}
\caption[]{Topological properties of the 1995 snapshot of the undirected WTW. 
{\textbf a)} Cumulative degree distribution $P(k)$ and comparison with a power law with exponent $-1.6$ (dashed line). {\textbf b)} Plot of $K_i^{nn}$ versus $k_i$. {\textbf c)} Plot of $C_i$ versus $k_i$.}
\label{fig2}
\end{figure}

One first-order topological property is the \emph{degree distribution} $P(k)$. In ref.\cite{mywtw} we showed that, contrary to previous results \cite{wtw} suggesting a power-law form of the distribution with exponent $-1.6$, the scale-free behaviour is suppressed by a sharp cut-off, which is actually restricted to a narrow region (see fig.\ref{fig2}a, where a power law with exponent $-1.6$ is shown as a comparison).

A second-order topological property, related to the form of the two-vertices joint degree distribution $P(k,k')$, is obtained by plotting the \emph{average nearest neighbour degree} (ANND), defined as ${K^{nn}_i=\frac{1}{k_i}\sum_j b_{ij}k_j}$, versus the degree $k_i$ (see fig.\ref{fig2}b). The decreasing trend shown in the figure clearly signals that the degrees of neighbouring vertices are anticorrelated, or in other words that the WTW is a \emph{disassortative} network \cite{assort}. From an economic point of view, this means that countries with many trade partners are on average connected to countries with few partners \cite{wtw,mywtw}.

Finally, the third-order correlation structure (related to the three-vertices joint degree distribution) can be investigated by plotting the \emph{clustering coefficient} ${C_i=\frac{1}{k_i(k_i-1)}\sum_{j\ne i}\sum_{k\ne i,j}b_{ij}b_{jk}b_{ki}}$ versus the degree $k_i$. As shown in fig.\ref{fig2}c, the trend is again decreasing, meaning that the partners of well connected countries are less interconnected than those of poorly connected ones \cite{wtw,mywtw}, a property sometimes referred to as \emph{hierarchy}.

\section{The \emph{fitness} network model and the role of the GDP}
A very intriguing result \cite{mywtw} is that the above topological properties of the WTW are in excellent agreement with the predictions of the \emph{fitness} network model \cite{fitness}, assuming that the probability that two vertices are connected is a function of some \emph{hidden variable} or \emph{fitness} associated to each of them (again, we are considering the WTW as an undirected network). 
In other words, if each vertex $i$ is assigned a fitness value $x_i$, the model assumes that the probability $p_{ij}$ that $i$ and $j$ are connected is a function $p(x_i,x_j)$ of $x_i$ and $x_j$ alone. All the expected topological properties therefore depend only on the functional form of $p(x,y)$ and on the fitness distribution $\rho(x)$. 
For the WTW, we showed that the \emph{Gross Domestic Product} (GDP in the following) of a country can be successfully identified with the fitness of the corresponding vertex \cite{mywtw}. Extending the formalism to the evolving case, 
if $w_i(t)$ denotes the GDP of country $i$ at time $t$ we can introduce the adimensional variable
\begin{equation}
x_i(t)\equiv\frac{w_i(t)}{\langle w\rangle(t)}=\frac{w_i(t)}{\sum_{j=1}^{N(t)}w_j(t)/N(t)}
\end{equation}
and interpret it as the \emph{fitness} of vertex $i$. Then, once the form of  $p(x,y)$ is chosen, all the expected topological properties predicted by the model can be compared with the empirical ones.
If at every timestep $t$ the GDP is regarded as (proportional to) the potential number of trade connections that a country can develop, then the natural choice for $p_t(x,y)$ reads \cite{park}
\begin{equation}\label{f}
p_t[x_i(t),x_j(t)]=\frac{\delta(t) x_i(t)x_j(t)}{1+\delta(t) x_i(t)x_j(t)}
\end{equation}
where $\delta(t)>0$ is the only free parameter of the model. The above form ensures that, at a given timestep, different realizations of the network with the same \emph{degree sequence} $\{k_i(t)\}_{i=1}^{N(t)}$ are equiprobable \cite{park}.
The parameter $\delta(t)$ can be fixed by requiring that the expected number of links in the undirected WTW
\begin{equation}\label{L}
\tilde{L}^u(t)=\frac{1}{2}\sum_{i=1}^{N(t)}\sum_{j\ne i}p_t[x_i(t),x_j(t)]
\end{equation}
equals the observed one $L^u(t)$. After that, the topology is completely specified: the expected degree distribution is obtained by computing the expected degree
\begin{equation}\label{k}
\tilde{k}_i(t)=\sum_{j\ne i}p_t[x_i(t),x_j(t)]
\end{equation}
while the expected average nearest neighbour degree is
\begin{equation}\label{ANND}
\tilde{K}^{nn}_i(t)=\frac{\sum_{j\ne i}\sum_{k\ne j}p_t[x_i(t),x_j(t)]p_t[x_j(t),x_k(t)]}{\tilde{k}_i(t)}
\end{equation}
and the expected clustering coefficient is
\begin{equation}\label{C}
\tilde{C}_i(t)=\frac{\sum_{j\ne i}\sum_{k\ne j,i}
p_t[x_i(t),x_j(t)]p_t[x_j(t),x_k(t)]p_t[x_k(t),x_i(t)]}{\tilde{k}_i(t)[\tilde{k}_i(t)-1]}
\end{equation}
For instance, for the year 1995 we found that $\delta(1995)=78.6$ in order to have $\tilde{L}^u(1995)=L^u(1995)$ \cite{mywtw}. The expected properties given by eqs.(\ref{k},\ref{ANND},\ref{C}) obtained with this choice are shown in fig.\ref{fig2} as solid lines superimposed to the empirical points. The accordance is excellent, indicating that the model indeed captures the basic aspects of the WTW topology. 
The comprehensive analysis of the whole data set that we are presenting here confirms that the same is true for each snapshot of the network. 
We can describe the time evolution in a compact way by studying the temporal change of the statistical GDP distribution and of the parameter $\delta(t)$ (see figs.\ref{fig3}a-b). 
More specifically, note that the discrete sums $\sum_i$ on the r.h.s. of eqs.(\ref{k},\ref{ANND},\ref{C}) can be replaced \cite{fitness} by integrals of the form $N(t)\int dx\rho(x,t)$ where $\rho(x,t)$ denotes the fitness distribution at time $t$. Therefore, besides $\delta(t)$, the model predictions only depend on $N(t)$ and $\rho(x,t)$ and the evolution of these three quantities completely specifies the evolution of the network (together with $r(t)$ if one is interested in the directed character of the WTW).
In fig.\ref{fig3}a we plot the fitness distribution in its cumulative form $\rho_>(x,t)=\int_x^\infty \rho(y,t)dy$. We find that the tail of the distribution always collapses to a time-independent Pareto tail with exponent -1, corresponding to a power-law $\rho(x,t)\approx\rho(x)\propto x^{-2}$.
The evolution of $\delta(t)$ during the 1950-1996 time interval is instead shown in fig.\ref{fig3}b. 

\section{Discussion and conclusions}
\begin{figure}[h]
\begin{center}
\includegraphics[width=.32\textwidth]{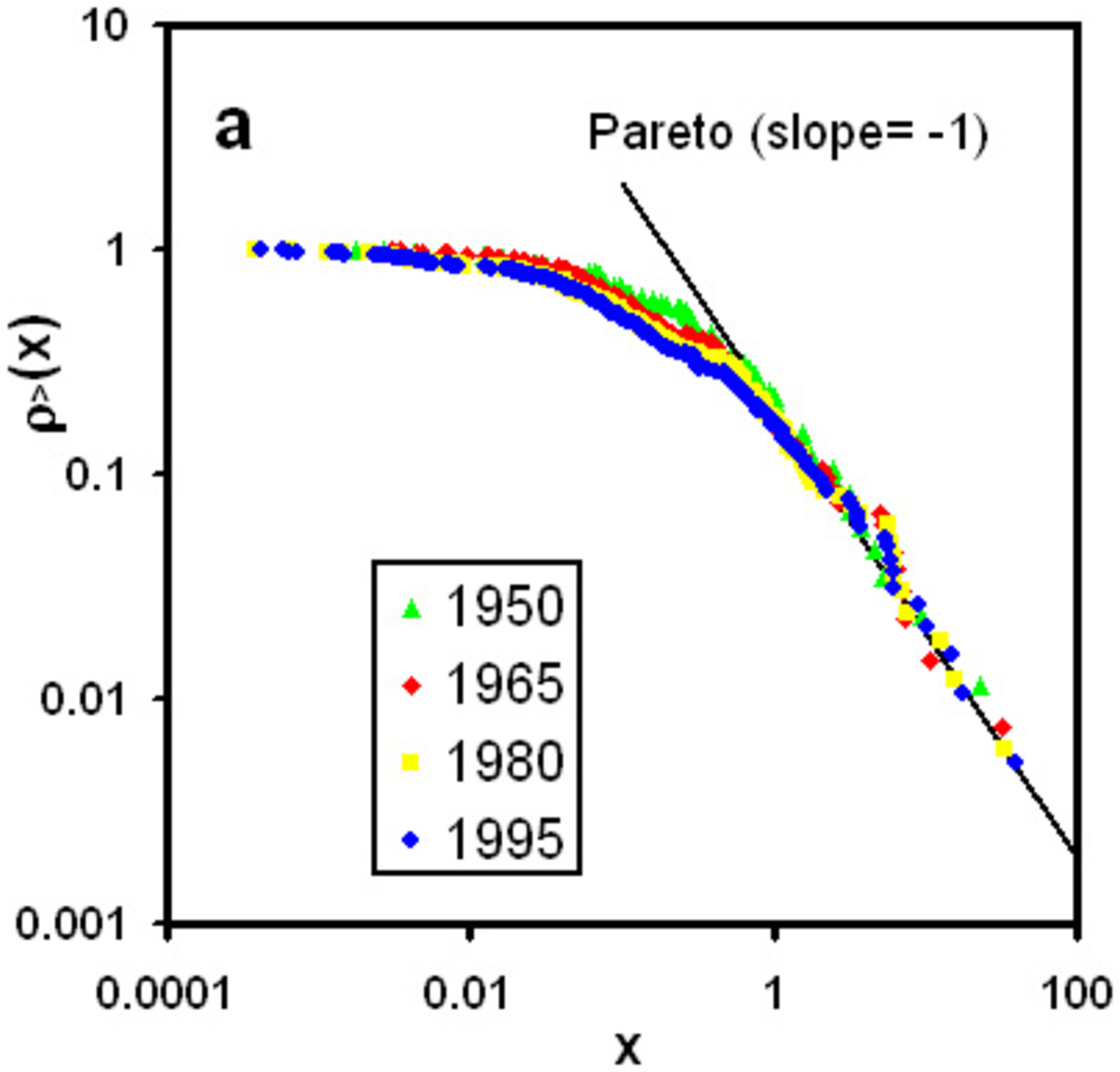}
\includegraphics[width=.33\textwidth]{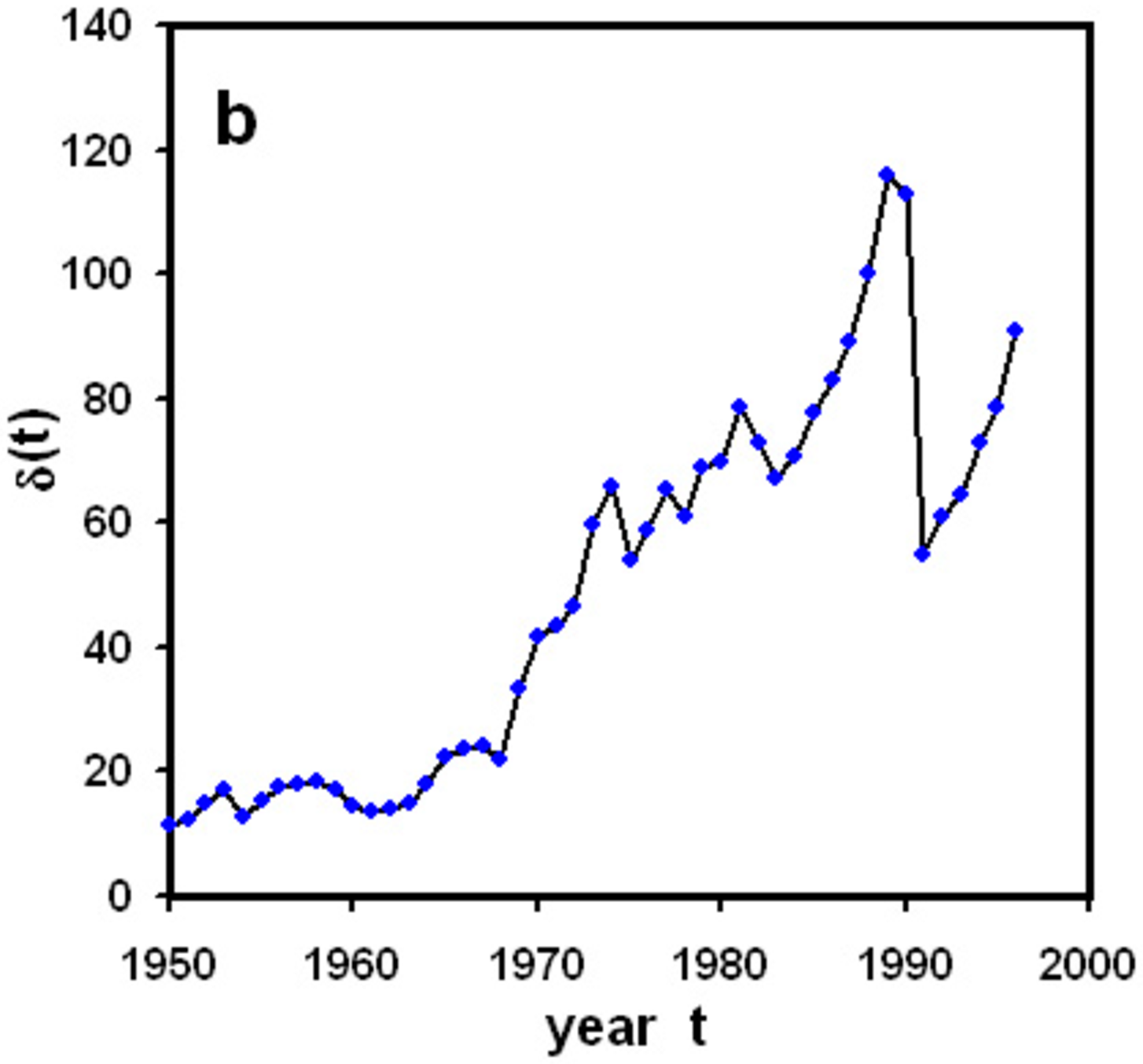}
\includegraphics[width=.33\textwidth]{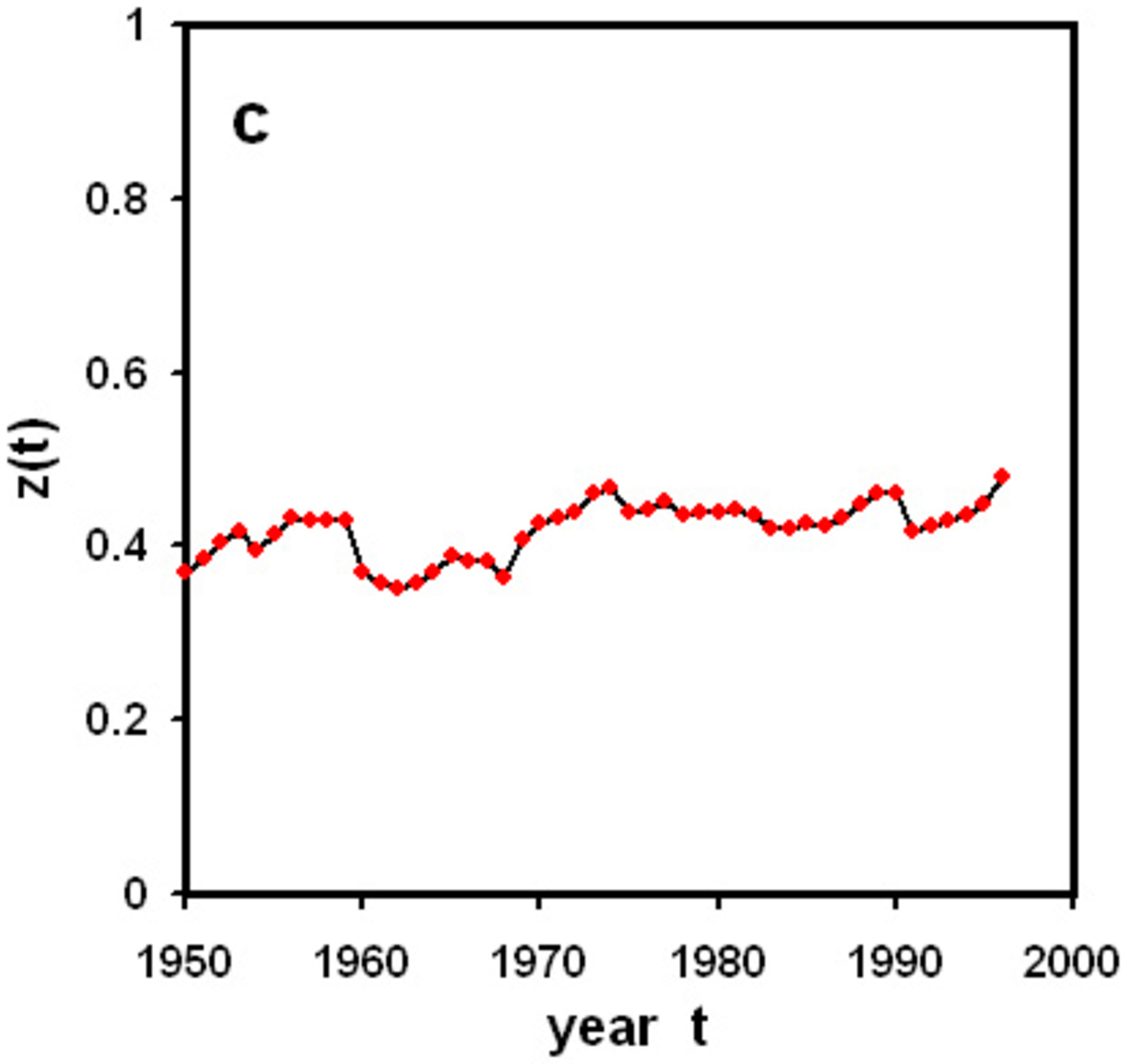}
\end{center}
\caption[]{{\textbf a)} Cumulative fitness distribution for four snapshots of the WTW (points) and comparison with a power-law with exponent -1 (solid line). {\textbf b)} Time dependence of the parameter $\delta(t)$ of the model. {\textbf c)} Time dependence of the link density $z(t)$.}
\label{fig3}
\end{figure}

A very important goal would be to relate the time variation of the parameters to the minimum possible number of external, unpredictable factors. For instance, the variation of $N(t)$ is of course due to complicated geopolitical reasons and should therefore be taken as an input information. By contrast, the seemingly irregular behaviour of $\delta(t)$ can be probably explained in terms of more regular quantities. For instance, we recall that $\delta(t)$ determines the number of links $L^u(t)=\frac{2-r(t)}{2}L(t)$ and is therefore related to the \emph{link density} $z(t)\equiv\frac{L(t)}{N(t)[N(t)-1]}$ (ratio of observed to possible links). Interestingly, we find that $z(t)$ displays relatively small fluctuations about the value 0.4 (see fig.\ref{fig3}c), a result related to the way new links form in the network (for instance, when a country `splits' in two or more countries, these can `inherit' the links of the initial country). 
The form and evolution of $\rho(x)$ has to be traced back to the dynamics of the GDP that takes place on the network. Models of wealth dynamics on complex networks \cite{BM,souma,dimatteo,mywealth} highlight how the topology determines the form of the wealth distribution. The results discussed here show that the reverse is also true, and therefore lead us to the intriguing picture of a countinuous feedback between dynamics and topology, a scenario that intimately relates the results of network theory to those of economic modelling.

\end{document}